\documentclass[12pt,preprint]{aastex}
\def\lax {\ifmmode{_<\atop^{\sim}}\else{${_<\atop^{\sim}}$}\fi}  
\def\gax {\ifmmode{_>\atop^{\sim}}\else{${_>\atop^{\sim}}$}\fi}  
\def\gtorder{\mathrel{\raise.3ex\hbox{$>$}\mkern-14mu
             \lower0.6ex\hbox{$\sim$}}}

\def\cm2{cm$^{-2}$}
\def\s1{s$^{-1}$}

\begin{document}

\title{Active galaxy nuclei: current state of the problem
}

\author{Elena Seifina\altaffilmark{1}}

\altaffiltext{1}{Moscow M.V.~Lomonosov State University/Sternberg Astronomical Institute, Universitetsky 
Prospect 13, Moscow, 119992, Russia; seif@sai.msu.ru}

\begin{abstract}
This review presents the main points of current advances in the field of active galactic nuclei (AGN).
A brief historical excursion about the search for the nature of AGN is given. The problem of close binary systems consisting of supermassive black holes located in the centers of galaxies is discussed in details. The main characteristics, as well as new methods for studying and ``weighing" these new objects, are described. This paper is based on a presentation made in the astrophysical seminar, which dedicated to the memory of the outstanding astrophysicist N.G.~Bochkarev (took place on May 19, 2023 at the Sternberg Astronomical Institute of Moscow State University).
\end{abstract}


\section{Introduction}
The activity of galaxies is usually associated with the variable brightness of their central parts (nuclei). Moreover, their variability, the so-called ``activity of galactic nuclei” (AGN), is accompanied by a powerful release of energy in a relatively small ($<$~1~pc) nuclear part of the galaxy, which is not explained by the activity of individual stars and gas-dust complexes located in them (Zasov \& Postnov, 2016). Despite intensive multi-wavelength studies of AGNs and significant advances in the study of the physical processes associated with them, their spatial structure, spectral features and mechanism of variability still remain not entirely clear. Currently, changes in the brightness of the central parts of galaxies are associated with the presence of a massive compact object, most likely a black hole (BH) in the galaxy center, which is the cause of the increased radiation intensity. However, this conclusion was not reached immediately. A long and winding path of research, starting in 1908, led to the discovery of these objects and their understanding as AGN.

\section{Brief history of AGN discovery}
\label{History} 
These objects were first discovered as ``spiral nebulae,” the spectra of which were distinguished by strong absorption lines on 
a continuous spectrum (Fath, 1908). These spectral features were perceived as errors and imperfections of the equipment, or unknown effects of spectrograms. However, after using an improved spectroscope, specially created by {\it Edward Fath} (1880 -- 1959) for faint objects, the problem did not disappear. And the only explanation could be the assumption that these ``nebulae” are star clusters that are not resolved into individual stars. This was one of the first proofs of the stellar nature of galaxies. In addition, observations of the ``spiral nebula” NGC~1068 showed that in addition to a continuous spectrum with absorption lines, bright emission lines, like those of ordinary gaseous nebulae, are clearly visible. Then these lines did not seem such an important discovery, but later they played a very important role in the study of AGN.

A few years later, this structure of the NGC~1068 spectrum was confirmed by {\it Vesto 
Slipher} (1875 -- 1969), and the absorption lines were identified with Fraunhofer lines. Slipher's further observations showed that the nebular nitrogen lines were emitted in a more compact region than the hydrogen lines. But the most important discovery was the unprecedented width of the emission lines. Having obtained a value of 300~km/s for the rotation velocity of the emitting region, Slifer rejected the Doppler interpretation of this broadening because the required velocities were too high. The ``enormous” broadening was immediately noticeable, since the lines in the spectrogram were no longer images of the slit, but became like disks.

\begin{figure}
\includegraphics[width=\textwidth]{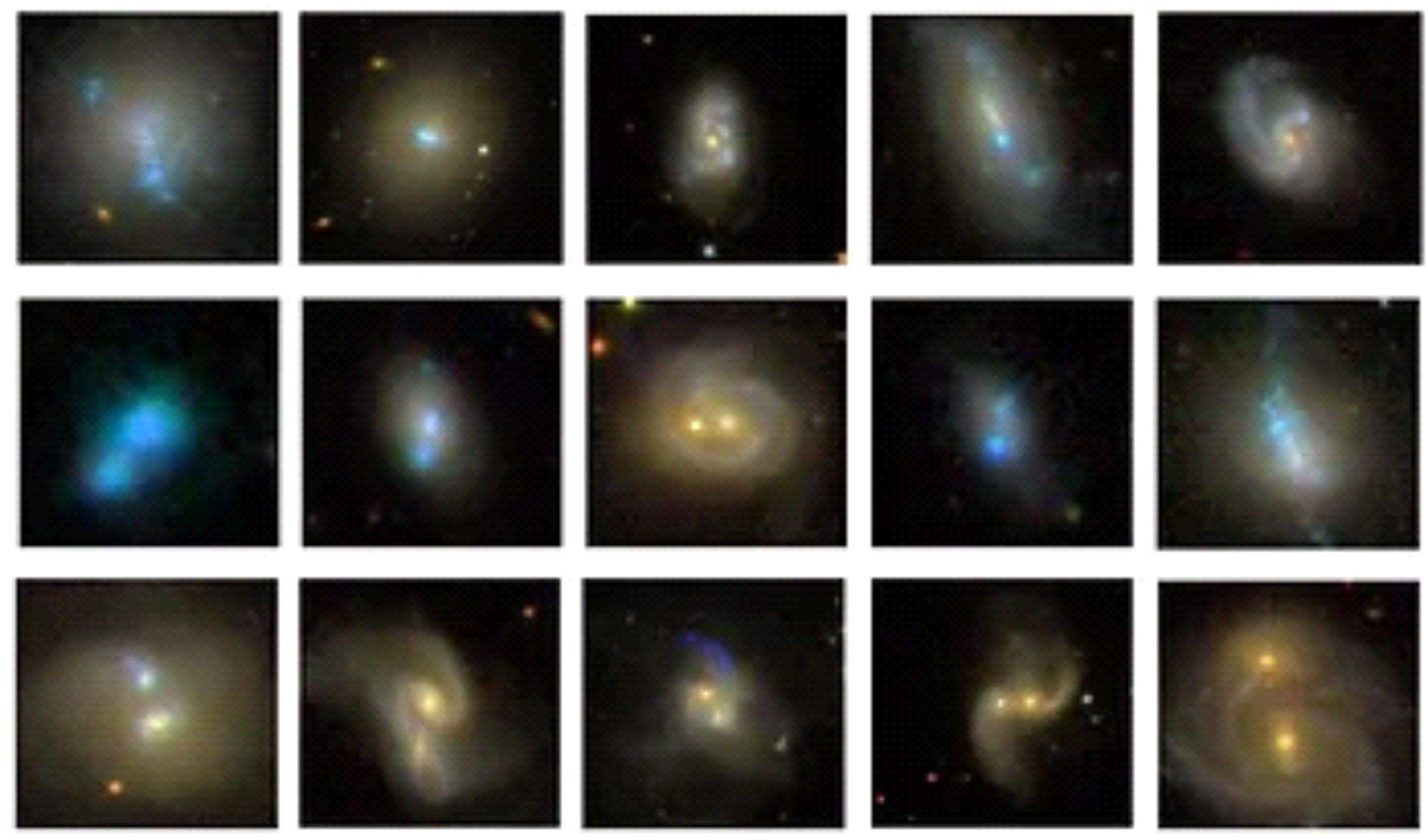}
\caption{{\it From top to bottom, left to right}: NGC~5058, NGC~3773, Mrk~1114, Mrk~712, Mrk~721, Mrk~116, Mrk~104, NGC~3758, Mrk~1263, NGC~7468, NGC~5860, Mrk~423, NGC~5256, Mrk~212, and MCG~+00-12-073. Figure is taken from Mezcua et al. (2014).
}\label{fig:example_2}
\end{figure}

Despite their intriguing spectral features, research of ``spiral nebulae" was continued only in 1943 by {\it Carl Seyfert} (1911 -- 1960). He discovered in the spectra of 12 galaxies wide (up to 8500~km/s) emission lines of hydrogen, helium and ionized iron, the half-width of which, in accordance with the Doppler effect, corresponded to velocities of up to several thousand km/sec (Seyfert, 1943). Later, galaxies with similar features were called {\bf Seyfert galaxies}. However, the study of the nature of these objects has almost stopped and has not aroused worldwide interest among astronomers. 
But unexpected discoveries came from where no one expected.

A decade before Seyfert's paper, {\it Karl 
Jansky} (1905 -- 1950) discovered increased radio noise of unknown origin in the meter wavelength range, while the direction towards the source of their maximum power made a full circle in 24 hours and approximately coincided with the direction towards the Sun. Jansky later identified the source of this radio noise with the central part of the Milky Way in 1933 (Jansky, 1933).

Astronomers almost did not notice Jansky’s discovery, but radio sky surveys gradually began, the coordinates of local maxima of the cosmic radiophone were clarified, and the upper limits on the size of objects became smaller and smaller. 
Thus, Hey et al. (1946) published a sky survey at 60 MHz, in which they identified a small source in the constellation Cygnus and determined that its diameter was no more than 2 degrees. In 1948, Bolton and Stanley 
reduced it to 8~arcminutes, and also determined the brightness temperature of the source to be $4\cdot 10^6$~K, which finally cast doubt on the thermal nature of the radiation. 
Finally, in 1954, Baade and Minkowski presented for the first time the optical identification of a radio source in Cygnus. These turned out to be two interacting galaxies (Baade \& Minkowsky, 1954), and the radio luminosity was higher than the optical luminosity. This is how the class of {\bf radio galaxies} was designated.

Further, by 1960, {\it Alan Sandage} (1926 -- 2010) and {\it Martin Schmidt} (1929 -- 2022) took up the optical identification of radio sources. Among other things, the object 3C~48 was identified, which in the optical range corresponded to a star with a very strange spectrum~--- it contained unknown and very broad emission lines. Such objects were called quasi-stellar radio sources 
or {\bf quasars}. 
And the first light on their nature was shed only in 1963, when Schmidt came to the conclusion that the unknown lines correspond to the Balmer series ($H_\beta$, $H_\gamma$, $H_\delta$ and $H_\epsilon$ lines) with redshift~$z=0.16$. This was confirmed by the prediction of the position of the $Mg II$ lines in the spectrum of the same quasar. 
Thus, the mystery of quasar spectra was solved.

\begin{figure}
\includegraphics[width=\textwidth]{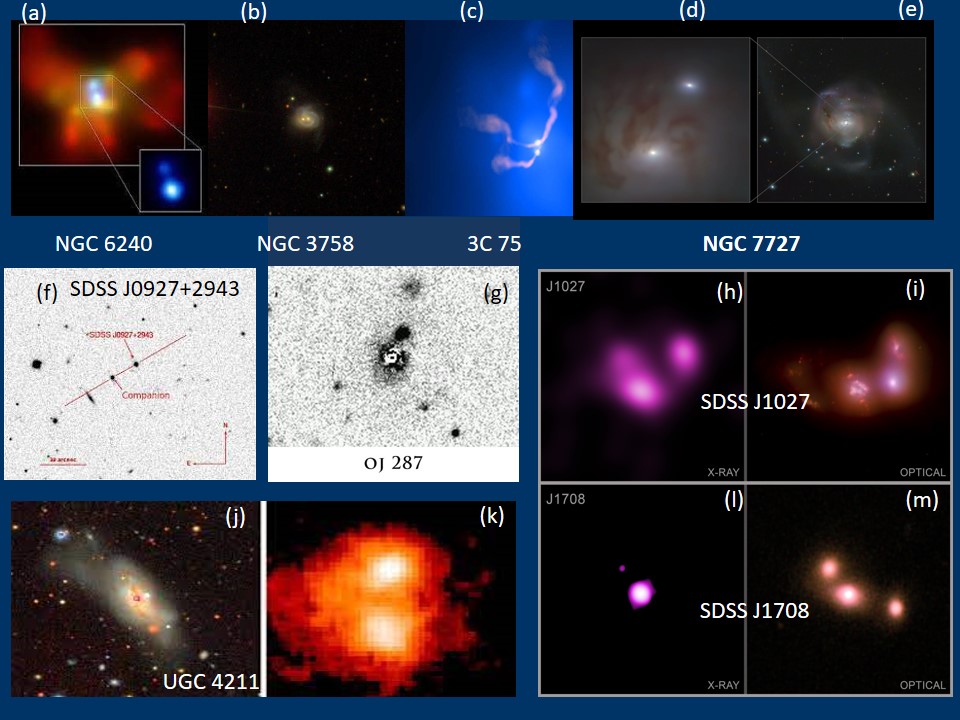}
\caption{Examples of AGNs containing binary black holes at their centers (top to bottom, left to right): NGC~6240 (a), NGC~3758 (b), 3C~75 (c), NGC~7727 (d, e), SDSS~J0927+2943 (f), OJ~287 (g), SDSS~J1027 (h, l),  SDSS~J1708 (i, m) and   UGC~4211 (j, k). 
}\label{fig:example_1}
\end{figure}

In fact, as possible reasons for such a redshift, Schmidt indicated only two plausible options: either gravitational redshift or cosmological. Preference was given to the cosmological redshift, resulting in a distance of 500~pc to 3C~273 at the Hubble constant 100 km/s/Mpc (Schmidt, 1965). The upper limit on the size of the emitting nuclear region was estimated at 1~kpc. At the same time, Matthews and Sandage guessed that the optical variability immediately implies the size of the emitting region, which for 3C~273 does not exceed 0.15~pc (Matthews \& Sandage, 1963). The extragalactic nature of quasars was soon confirmed (Greenstein \& Schmidt, 1964).

These, in brief, are the initial bold steps towards studying the nature of AGN. We have presented them for contrast and comparison with the level of understanding of AGN that has been achieved today.

\section{Current understanding of AGN activity}
\label{Current_AGN} 

In the present we are observing impetuous processes of evolution and outbursts of AGNs. The generally accepted model of AGN, which explains their increased radiation intensity, consists of a rotating massive central black hole and a surrounding accretion disk of gas, which is a source of powerful ionizing radiation. This model qualitatively explains the observed correlation of fluxes in the continuous spectrum and broad hydrogen lines, as well as the existence of a delay between them. The use of the reverberation mapping method to study the nuclei of Seyfert galaxies, in order to determine the structure of galactic nuclei and estimate the masses of supermassive black holes in the centers of galaxies, was first proposed by {\it Nikolai Gennadiavich Bochkarev} (1947 -- 2022). Long-term spectral monitoring of AGN, together with the development of the reverberation analysis method (Wanders et al., 1997; Bochkarev \& Gaskell, 2009), suggested that the emission of broad hydrogen emission lines arises in gas clouds moving along Keplerian orbits in approximately the same plane and forming an outer disk. But there is no general agreement among experts on this matter yet. Recently, in world research, special attention has been paid to studying the relationship between AGN radiation in the X-ray, optical and radio bands. Despite some progress achieved in the study of AGNs, many problems and tasks remain unresolved, for example, the manifestations of binary black holes (BBH) in the nuclei of galaxies.

\begin{table*}
\scriptsize
  \centering
  \caption{Examples of active galaxies 
hosting binary/triple black holes at their centers. }
\label{tab1}
\begin{tabular}{l c c c  l r}
\hline
Source name & Distance,   & z  & Class  & Duality   & M$_{BH}^{\dagger\dagger}$, \\
                    & Mpc         &     &          & criterion$^{\dagger}$ &  $\times$10$^7$M$_{\odot}$\\
\hline
NGC 7727$^{(1,2,3)}$ & 27.4  & 0.006 & SAB(s)a pec & Im &  15.4, 0.6 \\
NGC 6240$^{(4)}$ & 122.6      & 0.024 & E              & Im &  $\sim$36, 71, 9 \\
NGC 3758$^{(5)}$& 130.3      & 0.029 & Sab & Im &  $\sim$ 1, $M_x^{\dagger\dagger\dagger}$ \\
UGC 4211$^{(6)}$ & 153         & 0.035 & Sy 2/S0-a & Im, Sp &  $\sim$ 10, 10 \\
SDSSJ0927+2943$^{(7)}$ & 2860    & 0.713 & QSO       & Im, Sp &  14, $M_x^{\dagger\dagger\dagger}$ \\
SDSSJ1537+0441$^{(8)}$ & 1250   & 0.380  & QSO       & Sp                  & 2, 80 \\
3C 75$^{(9)}$           & 92      & 0.023   & E          & Jet            &  $>10^{4}$, $M_x^{\dagger\dagger\dagger}$ \\
OJ 287$^{(10)}$        & 1037  & 0.306   & BL Lac    & Lc          &  12.5, 2$\times 10^{3}$ \\
\hline 
\end{tabular}
\\
{\footnotesize {\bf Notes.} $^{\dagger}$ Im -- image, Sp -- spectrum, Jet -- similar bend of numerous jets 
Lc -- periodic double-peak light curve;  
$^{\dagger\dagger} M^i_{BH}$ masses of the components of the BH system, $i=1,2,3$ is the numbers of components for binary or triple BH system (where applicable); 
$^{\dagger\dagger\dagger}$ case when the BH-companion mass is unknown. \\
{\bf References.}  
$^{(1)}$Tully et al.(2016); 
$^{(2)}$NASA/IPAC Extragalactic Database;
$^{(3)}$Voggel et al.(2022);
$^{(4)}$Kollatschny et al.(2020); 
$^{(5)}$Koss et al.(2011); 
$^{(6)}$Koss et al.(2022); 
$^{(7)}$Decarli at al.(2010); 
$^{(8)}$Boroson \& Lauer(2009); 
$^{(9)}$Hudson et al.(2006), Paltani \& Turler(2005); 
$^{(10)}$Titarchuk et al.(2023).}
\end{table*}

These new objects combine the properties of objects of different classes and scales: {\it small-scale} binary systems (BSs) and  {\it large-scale} AGNs. Moreover, the discovery of a large number of such objects simultaneously in different wave bands significantly contributes to such a synthesis of BS and AGN problems. With the advent of the X-ray era, this new direction became quite promising and represents a new actual problem in the study of AGNs. 

\section{The dawn of the era of binary black hole research}
\label{BBH} 

Numerous reports of the discovery of binary black holes in the centers of galaxies have been pouring out like a cornucopia since the end of 2021. To date, their number already amounts to more than a hundred reliably established binary black holes. With the accumulation of such a number of BBH objects, a turning point came when it was no longer possible to ignore such objects. And an urgent need arose for their study, including the urgent question of weighing such black holes and their origin  in general. The problem is that classical methods, for example, the ``dynamic” method (Karttunen et al., 2017; Martynov, 1988) no longer work due to the significant distance of the objects and, as a consequence, the impossibility of resolving/separating the radiation of components. Furthremore, BHs do not emit in any band of electromagnetic radiation, and we can judge their presence only by their effect on the surrounding matter. But how to divide this influence in the BBH case? Therefore, new approaches to assessing the duality of BHs and improving methods for studying each of  two BHs separately are relevant.

Thus, classical methods of stellar kinematics and dynamics are more likely to give an estimate of the total mass of two BHs. The ``dynamic'' method is universal and is considered to be the most accurate method for ``weighing” of BS components today (Cherepashchuk, 2023), but it requires measuring the signal (radial velocity) from each of the BS components separately and requires knowledge of many system parameters that are not known a priori and must be obtained by indirect methods. Recently, a new method was proposed for estimating the mass of the ``central engine” in the case of a BBH in the center of the galaxy from X-ray observations -- the so-called ``scaling” method (Titarchuk et al., 2023).

The study of binary systems hosting two SMBHs has become a separate, very important field of relativistic astrophysics. This raises many questions: what properties do binary black holes have, and how much does their behavior differ from single black holes? In particular, is the activity of such galactic nuclei a new type of AGN activity? Or can they be attributed/associated with some subtype of already known ones (Seyfert, quasars, blazars, radio galaxies)? This is what this paper will discuss. At the same time, of course, the paper goal is not to tell everything and exhaust the topic. Rather, with the help of individual strokes, create an image of those problems that have been solved or are being solved in the field of AGNs.

The physics of galaxies with central BBHs is the cutting edge of astrophysics, where many questions naturally arise related to scenarios for the formation of BBH systems, the stability and evolution of such systems. Nobody knows the final answer yet. But we may benefit from the experience of studying both BS and AGN, as well as the analysis of numerous observations of binary SMBHs.

\section{Binary systems and AGNs: synthesis of problems}
\label{CBS+AGN_synthesis} 

First, a few words about the classical definition of a stellar binary system 
and a close BS (CBS). As is known, a BS is a system of two gravitationally bound stars revolving in closed orbits around a common center of mass. In turn, a close binary system is a subtype of a binary binary system in the case when  these stars can exchange mass with each other (Cherepashchuk, 2023). By now, the concepts of BS and CBS have received an unexpected extension, which consists in the fact that the components of the BS and CBS can be not only stars, they can also be black holes. We are already well know about X-ray BS, in which one of the components is a compact object, for example, a stellar-mass BH (3--10 M$_{\odot}$). However, the update of the BS concept is that a BS can contain two BHs at once, moreover, two supermassive BHs (SMBHs, $>10^5$ M$_{\odot}$), which rotate in closed orbits around a common center of mass and with mass exchange. 
The exchange of mass between black holes in such a binary system sounds strange, because it is known that even light, let alone matter, cannot escape from under the event horizon of a black hole.
If this is really the case, then mass exchange in CBSs hosting SMBHs can be put on a par with the evaporation of black holes according to Hawking. This is why the study of just such binary systems is interesting.

It should be noted that the idea of binary systems with supermassive black holes was first proposed by Boris Valentinovich Komberg back in 1967. Subsequently, his hypothesis about the possible duality of nuclei in quasars (Komberg, 1967) was brilliantly confirmed in observations (Sect.~\ref{Observational_signatures}).

However, the method of establishing the duality of BHs is a complex and responsible task, especially in the case of binary systems with supermassive BHs. Now this direction is just being formed and no one knows the final answer yet. As a result of long-term work by theorists and observers, five observationally most suitable methods of signs have been identified, which we list below and illustrate their application with examples of such systems (see Table~\ref{tab1}), which today are the most reliable candidates for binary black holes.

\section{Observational manifestations of the duality of galactic nuclei}
\label{Observational_signatures} 

Where did the suspicion about the duality of certain galactic nuclei come from? First of all, based on their observational images for {\it nearby} galaxies, which make it possible to see the structure of their central regions using telescopes with good resolution (see Sect.~\ref{duality_image}, Figure~\ref{fig:example_2}). In addition, spectroscopic studies make it possible to establish the duality of {\it distant} AGNs from the characteristic behavior of spectral line systems (see Sect.~\ref{duality_spectrum}). Scientists have long suspected that when two galaxies collide, their nuclei (with one SMBH in the nucleus of each galaxy) do not merge together immediately, but first form a binary system of these SMBHs (Figure~\ref{fig:example_1}). 
Some observational indication/confirmation of this guess for specific galaxies, the parameters of which are listed in Table~\ref{tab1}, will be presented below.



\begin{figure}
\centering
\includegraphics[scale=0.70,angle=0]{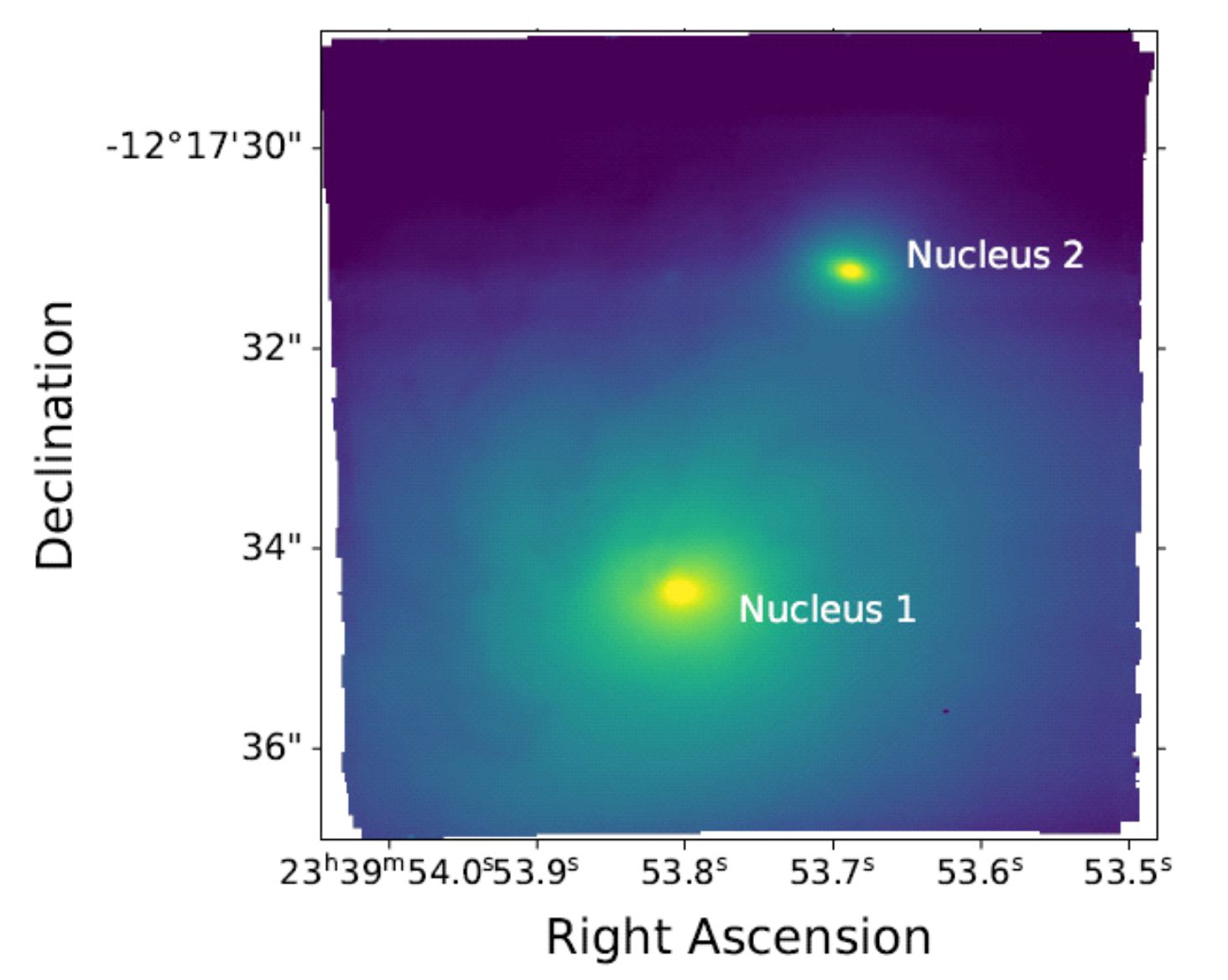}
\caption{Active galactic ``Nucleus 1" is the photometric center of the ``main" galaxy NGC~7727, and active galactic ``Nucleus 2" lies to the northwest, with an expected distance of 1600 light years. Image was taken from Voggel et al. (2021).
}
\label{fig:example_0}
\end{figure}

\subsection{SMBH duality in AGN, established by its image}
\label{duality_image} 

\subsubsection{NGC 7727}
\label{NGC 7727}

The first candidate for a binary-BH AGN is the {\it nearby} galaxy NGC~7727, which is located at a distance of only 27.4 Mpc. In the image taken in 2021 with the Very Large Telescope's multi-unit spectroscopic explorer (MUSE) at the European Southern Observatory, the two bright nuclei are clearly visible  in the wide image (Fig.~\ref{fig:example_1}d) and compact image (Figure~\ref{fig:example_1}e). Special MUSE spectrometric equipment has made it possible to clearly identify the active galactic nucleus 1 (``Nucleus 1” in Figure~\ref{fig:example_0}), which is the photometric center of the ``main” galaxy NGC~7727, and active galactic nucleus 2 (``Nucleus 2” in Figure~\ref{fig:example_0}), which lies to northwest at an expected distance of 500~pc. Karina Voggel and her colleagues (Vogel et al., 2022) showed that a larger BH is located in the center of the galaxy NGC~7727, while the smaller BH is its satellite. Over time, they will merge into one giant black hole. The masses of the two objects were determined by observing how the gravitational attraction of black holes affects the motion of the stars around them. According to their calculations, the mass of the larger BH of the pair is $1.5\times 10^8$ M$_{\odot}$, and the mass of the smaller BH is $6.3\times 10^6$ M$_{\odot}$. 


\begin{figure}
\centering
\includegraphics[width=\textwidth]{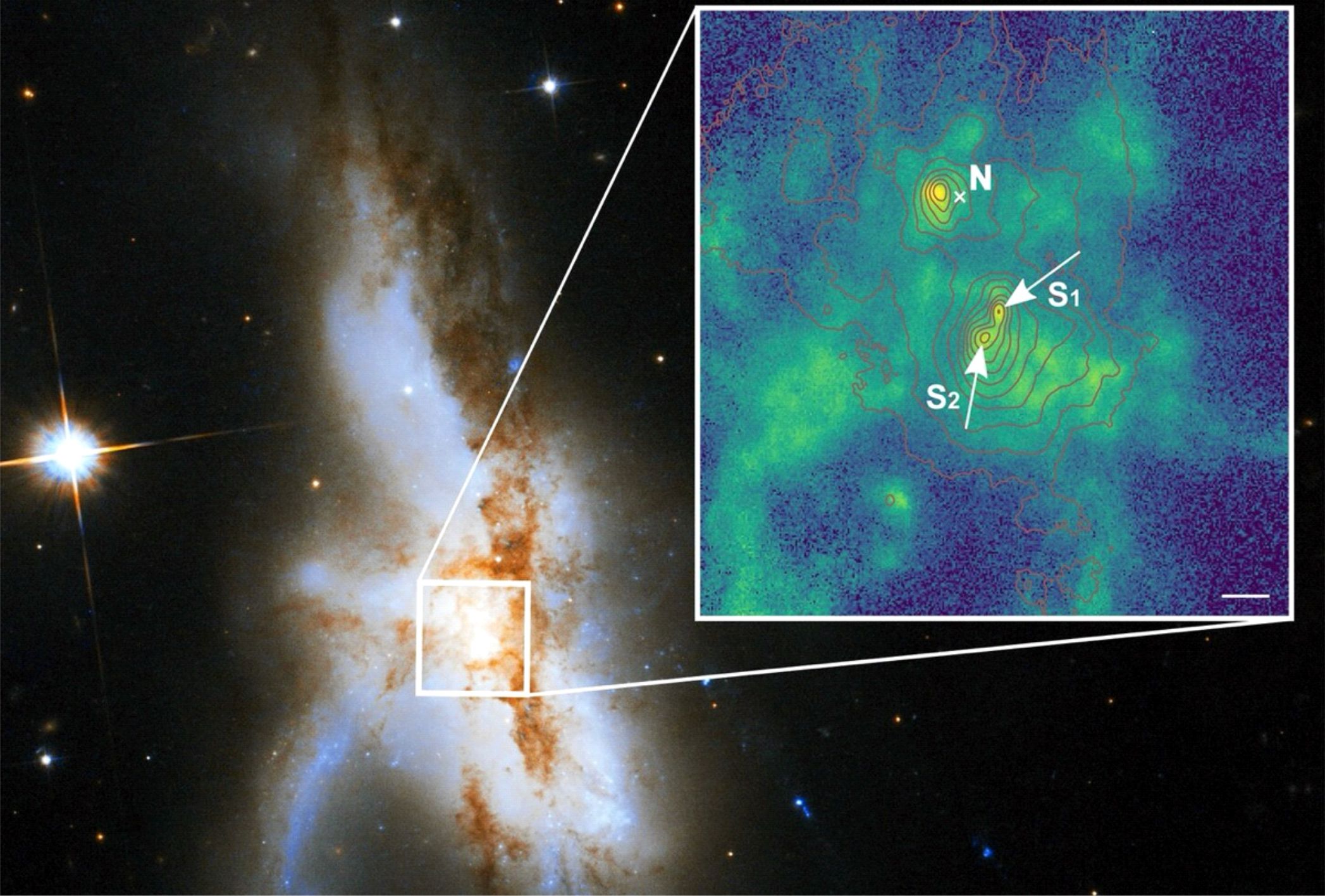}
\caption{Image of the galaxy NGC~6249 using HST (left) and an enlarged image of the galaxy central region in the zooming box (top right) using the MUSE instrument in the Narrow-Field Mode with the four-laser GALACSI adaptive optics system in the ESO VLT. MUSE made it possible to detect a ``northern” ($N$) SMBH and two ``southern” ($S_1$ and $S_2$) SMBHs. Green in the plot shows the gas distribution H$_{\alpha}$/[NII]. Red overlaid are $I$ band contour levels. Image was taken from Kollatschny et al. (2020).
}
\label{fig:example_4}
\end{figure}

\subsubsection{NGC 6240}
\label{NGC_6240} 
Next, the X-ray image of NGC 6240 obtained with {\it Chandra} (shown in blue in bottom right box), superimposed on the optical image of the galaxy (shown in red in top left panel, Figure~\ref{fig:example_1}a), clearly shows two sources at a distance of about 900~pc (Kollatschny et al., 2020) between them (projected onto the sky plane). Interestingly, the presence of a  double active nucleus -- two black holes -- at the center of NGC~6240 was known back in 1983. Since then, numerous studies of the galaxy have been carried out. But because the black holes are so close to each other, their exact locations have remained elusive. Later, thanks to observations with extremely high spatial resolution, it was established (Kollatschny et al., 2020) that one of them, the southern one, is in fact itself a binary system of supermassive black holes (Figure~\ref{fig:example_4}). Each of the supermassive black holes has a mass exceeding $9\times 10^7$ M$_{\odot}$ (see Table~\ref{tab1}). Moreover, these three objects are at such a distance that we can say that the system is at a fairly late stage of its merger, which will take more than a billion years. This is the closest multiple supermassive BH system to Earth. 

\subsubsection{NGC 3758 (Mrk 739)}
\label{NGC 3758} 
Another similar nearby galaxy containing two active SMBHs separated by a distance of only 11 thousand light years is NGC~3758 (Figure~\ref{fig:example_1}b). This is a peculiar spiral galaxy that is located at a distance of 130 Mps. The discovery of the duality of this galaxy's nucleus in X-rays was made possible by {\it Swift} monitoring in the search for cosmic sources of high-energy X-ray radiation. In addition, the capabilities of the {\it Chandra} X-ray telescope, which does not provide such a wide view, but a couple of orders of magnitude higher resolution, were also used. The duality of the central part of the galaxy NGC~3758 itself has been known for quite some time (Figure~\ref{fig:example_1}b), but the duality of its nucleus was confirmed by an X-ray image in which two bright centers are clearly visible. Koss et al. (2011) indicates  the mass of one of the BHs of this binary system as 10$^7$ M$_{\odot}$. The mass of the second black hole has yet to be measured. 

\subsubsection{UGC 4211 }
\label{UGC 4211 } 
Galaxy UGC~4211 (Figure~\ref{fig:example_1}j) with two nuclei clearly visible in the enlarged image (Figure~\ref{fig:example_1}k) is located at a distance of $\sim$470 million light years. The black holes have similar masses, $\sim 10^8$ M$_{\odot}$, and are located at a distance of 230~pc from each other, making UGC~4211 as the galaxy with the shortest spatial separation of the binary nucleus components (Koss et al., 2022). 

\subsection{SMBH duality in AGN, established by spectrum analysis}
\label{duality_spectrum} 

\subsubsection{SDSS J1537+0441}
\label{SDSS J1537+0441} 
Quasar SDSS~J1537+0441 is located at a distance of 4.1 billion light years. Presumably, the quasar hosts two BHs orbiting 9.5$\times$10$^{12}$ km from each other. The spectrum of SDSS~J1537+0441 has two systems of broad lines, shifted relative to each other in the spectrum by a distance that corresponds to a relative velocity of 3500~km/s. In addition, the source spectrum shows a third system of unresolved narrow absorption lines that has an intermediate shift relative to the  first and second systems of broad lines. Everything looks as if in the center of  the galaxy, within a single region of narrow lines several light years in size, two black holes are moving around a common center of mass, each with its own region of wide lines. 

Estimates of the mass of these two black holes are made based on the size of the broad H$_{\beta}$ line: $2\times 10^{7.3}$ and $8\times 10^{8.9}$ M$_{\odot}$ (Boroson \& Lauer, 2009). These SMBHs orbit with a period of $\sim$100~years around each other with a distance of $\sim$0.1~pc between them.

Particular interest to the BH binary system in SDSS~J1537+0441 is caused by its very interesting stage of evolution -- a kind of ``dead zone” of orbital evolution. These BHs are already close enough to each other that there are not enough stars around them to ensure further convergence of the BHs due to dynamic friction. At the same time, they are still too far away to lose a significant energy and get closer due to the emission of gravitational waves. 

\subsubsection{SDSS J0927+294}
\label{SDSS J0927+294} 
SDSS~J0927+2943 is an unusual quasar (Figure~\ref{fig:example_1}f) whose spectrum shows two sets of optical emission lines at different redshifts. The first, or ``red” set consists of narrow [O~III] lines and other ionized elements with a redshift of $z$=0.712. Such lines are usually associated with gas far from the center of the galaxy, excited by radiation of the central quasar. The second, ``blue" set consists of broad emission lines that are typically associated with hot gas very close to the galaxy center  (and presumably to the central SMBH);  
these lines have a redshift of 0.697. The velocity difference between the two emission line systems is 2650~km/s. The origin of these 
line systems is believed to be due to gravitational wave recoil: the ejection of a supermassive black hole from the center of the host galaxy. In this interpretation, one set of emission lines arises from gas associated with the SMBH, 
and the other set is associated with gas left behind in the galaxy. Precisely, when a black hole is ejected from the nucleus of a galaxy, this black hole carries away gas emitting broad lines, leaving behind most of the gas emitting narrow lines. Such an ejection is possible if the black hole was recently formed as a result of the merger of two smaller black holes. The merger is accompanied by the emission of gravitational waves, which can be emitted anisotropically, imparting linear momentum to the merger remnant -- essentially ``knocking” the black hole out of the galaxy.

Decarli at el. (2010) estimated the mass of the central SMBH in SDSS J0927+2943 in 14$\times 10^7$ M$_{\odot}$. This discovery is important because it indirectly proves that black holes merge and that mergers can be accompanied by strong emissions. This process has been postulated by theory, but has never previously been confirmed by direct observation.

%
%
\begin{figure}
\centering
\includegraphics[width=\textwidth]{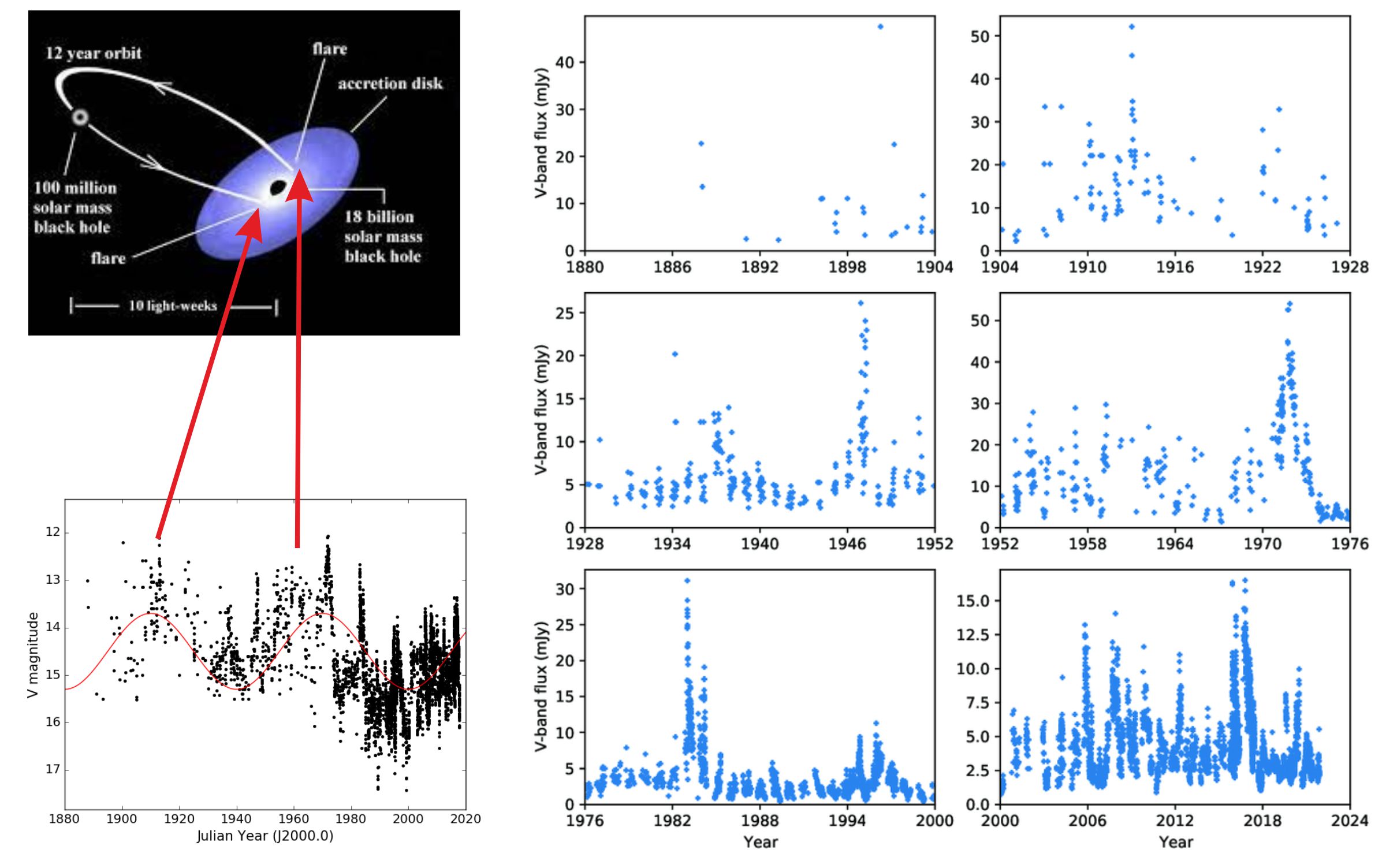}
\caption{To the discussion of the duality of the object OJ 287 [red arrows show points on the disk around the ``primary” BH, at which it is crossed by the orbit of the “secondary” BH near the periapsis; orbital eccentricity $e =$ 0.65 (Valtonen et al., 2012)]. Bottom left: Double peak of the optical light curve as it passes periapsis. Right: Optical light curve of OJ~287 (1888 -- 2021) with obvious periodicity.
}\label{fig:example_3}
\end{figure}

Thus, as shown above,  systems of two SMBHs in AGN centers have already been repeatedly discovered due to the fact that they are located close to the Earth (27 -- 130~Mpc) and their duality has been proven from images or spectral features. It is clear that for more distant sources the above methods are no longer applicable. In fact, the collision of such galaxies (and, accordingly, their nuclei) in the past, which led to the formation of a binary system of SMBHs, does not in any way demonstrate the duality of their new ``nucleus”, except perhaps for the presence of ``extra” jets (Sect.~\ref{duality_jet}) or strict periodicity outbursts  from this object (Sect.~\ref{duality_periodicity}).

And in such cases, the duality of the nucleus can sometimes be established by individual exotic/peculiar features, for example, by the same bending of numerous jets from the central part of  AGN (presumably from each of the SMBHs), by the specific shape of the light curve (for example, the periodicity and two-humpedness of the burst maximum on the light curve) or as a possible way to connect the results on optical, X-ray and radio data for the same object. Below we present such cases.

\subsection{SMBH duality in AGN, established from  peculiar jet shape (3C~75)}
\label{duality_jet} 
\label{3C 75} 
3C~75 is a system of two SMBHs in the galaxy NGC~1128, which is part of the Abell~400 galaxy cluster, 300 million light-years away. In the 3C~73 image (see Figure~\ref{fig:example_2}c), summing the X-ray (shown in blue) and radio (shown in pink) bands, two bright sources in the center are clearly visible -- these are two SMBHs orbiting around a common center of mass, feeding the powerful radio source 3C~75. 
The distance between these SMBHs is $\sim$7.6 kpc. 
These BHs are believed to be located in the ``new" nucleus of two merging galaxies. Hudson et al. (2006)  concluded that two supermassive black holes are connected by gravitational forces and form a binary system. They also made this conclusion for another reason. The jets ejected by the SMBHs have the same bend, which is most likely due to the fact that both SMBHs are rushing through the hot gas of the cluster at a velocity of 1200~km/s. The mass of one of the SMBHs of this pair is estimated $\ge 10^{13}$ M$_{\odot}$ Hudson et al. (2006).

%
%
\begin{figure*}
\centering
\includegraphics[scale=0.50,angle=0]{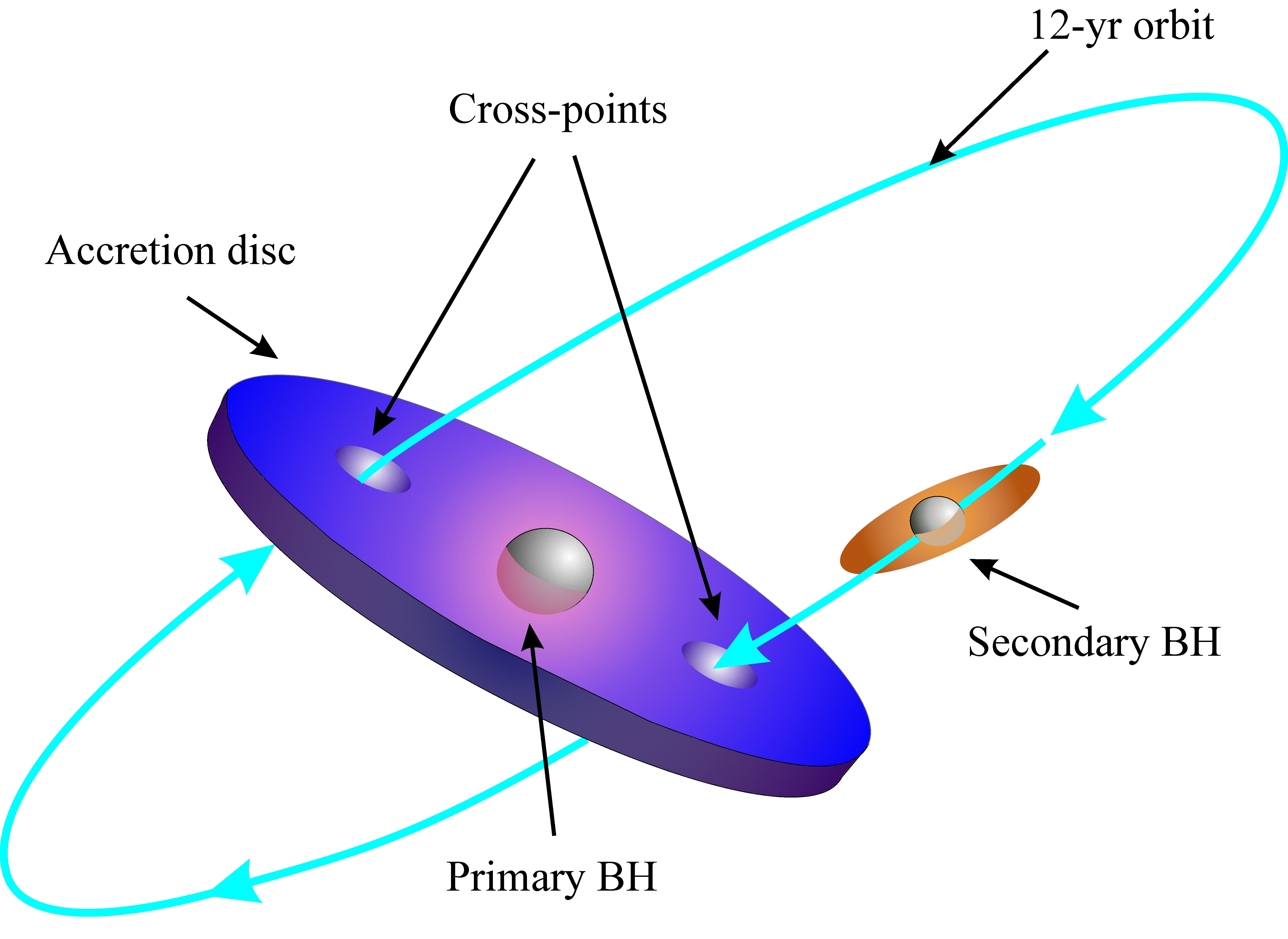}
\caption{
Schematic representation of OJ~287 binary system of supermassive black holes forming an orbital pair located in the OJ 287 galaxy. The ``primary" black hole is surrounded by an accretion disk. The ``secondary” BH ($\sim 10^8 M_{\odot}$) rotates around the ``primary” BH ($\sim 10^{10} M_{\odot}$), piercing its accretion disk twice every 12 years. When a ``secondary" BH passes through the disk around a ``primary" BH, a flare occurs in the X-ray, optical and radio ranges.
}
\label{picture}
\end{figure*}

\subsection{SMBH duality in AGN, established from  the periodicity of activity outbursts (OJ~287)}
\label{duality_periodicity} 

One striking example of such systems is a pair of SMBHs in the center of the galaxy OJ~287. However, for a long time it did not arouse much interest and its flare activity was interpreted only as known AGN activity, which is typically observed in 10\% of all galaxies. But usually the variability of AGN is characterized by spontaneity of outbursts and a clear absence of strict periodicity. In contrast, outbursts from the nuclear part of the galaxy OJ 287 occurred with stunning periodicity -- with a period of 12 years (Figure~\ref{fig:example_3}, right). Interestingly, these outbursts appear at different wavelengths -- from radio to X-rays. The latest of these multi-wavelength outbursts in OJ~287 was observed by the X-ray telescope (XRT) on board the Neil Gehrels $Swift$ Observatory and has generated much debate about whether the source contains one or two black holes and what the characteristics of this binary system are. It was the strict periodicity of brightening observed in OJ287 for more than a hundred years, starting in 1888, that helped to understand the nature of the activity of the nucleus of this galaxy, 
as a periodic orbit of the components of this huge binary system. In addition, the duality 
of the object OJ~287 is indicated by a specific detail in the light curve of OJ ~287, namely, its double peak (Figure~\ref{fig:example_3}, bottom left), which immediately indicates the discrepancy between the orbit of the small SMBH and the plane of the accretion disk around the heavy SMBH (Figure~\ref{fig:example_3}, top left).

 \begin{figure*}
 \centering
\includegraphics[width=13cm]{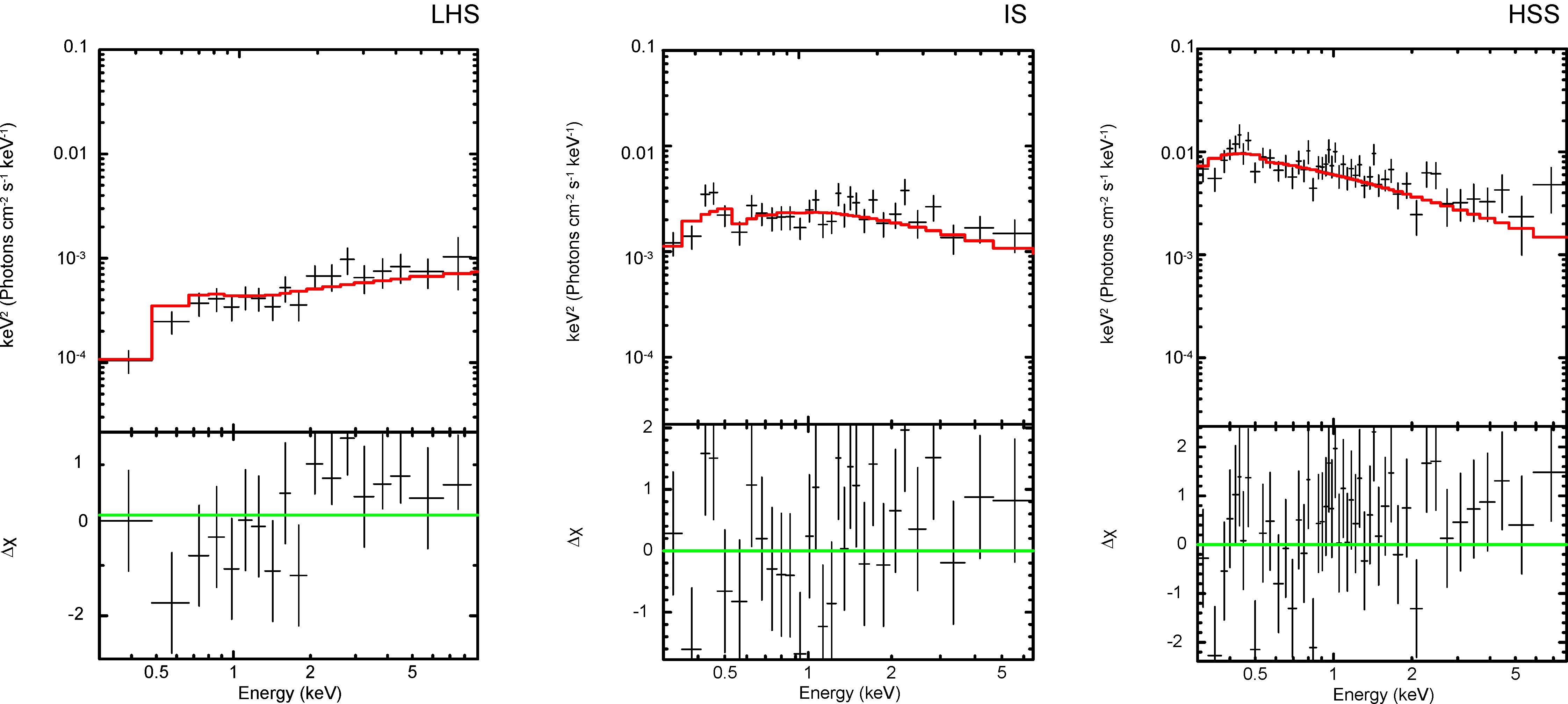}
   \caption{Three representative spectra of OJ~287 from {\it Swift}/XRT data with best-fit modelling for the 
LHS ($id$=00035011001), 
IS ($id$=00088085001)  and 
HSS ($id$=00034934051) states in units $E*F(E)$ using  the {\it Bulk Comptonization model} ($BMC$), modified by interstellar absorption ({\tt tbabs*bmc}). 
  The data are denoted by $black$ {crosses}, while the spectral model presented    by $red$ histogram. Figure is taken from Titarchuk et.al. (2023).
}
\label{3_spectra}
\end{figure*}
 
This phenomenon has been seen before, but on a more compact scale -- in binary star systems, and they were located closer to Earth, which made it possible to see them with the help of telescopes. At a great distance there is no such possibility, and it is possible to determine whether the galactic nucleus is a binary system of SMBHs only by indirect signs -- for example, by the nature of the outbursts. This raises some difficulties in interpreting the observations obtained.

The outburst occurs when a ``smaller" black hole passes through the disk of a galaxy. In this case, a process similar to the tidal destruction of stars occurs. A ``smaller" black hole(denoted in Figure~\ref{picture} as ``secondary” BH)  absorbs matter from the galactic disk near the periastron passage point, this matter heats up and emits photons in a wide range of wavelengths -- from radio to X-rays. Thus, a ``smaller" black hole reveals itself to an Earth observer as an outburst. From the nature of this outburst it can also be understood that the orbit of the secondary black hole does not coincide with the ``larger" BH accretion disk.


Titarchuk et al. (2023) determined the mass of the smaller SMBH in the galaxy OJ~287 by analyzing the source spectrum during  the outburst activity detected from 2006 to 2018 using XRT/$Swift$. They noticed that the spectrum shape  evolves in a manner characteristic of outbursts from black holes (Figure~\ref{3_spectra}) during BH transition between {\it low-hard state} (LHS) through {\it intermediate state} (IS) to {\it high soft state} (HSS). Namely, a monotonous increase in the photon index $\Gamma$ was discovered as the brightness of the object increased (the beginning of the outburst) and the index $\Gamma$ reached a ``plateau” (or ``saturation”) during the maximum of the X-ray outburst. In this case, the source radiation was analyzed in first principles models taking into account the Comptonization of accretion disk (AD) photons around the BH with hot electrons from the AD inner layers and the relativistic converging flow onto the BH.

%
%
 \begin{figure*}
 \centering
 \includegraphics[width=10cm]{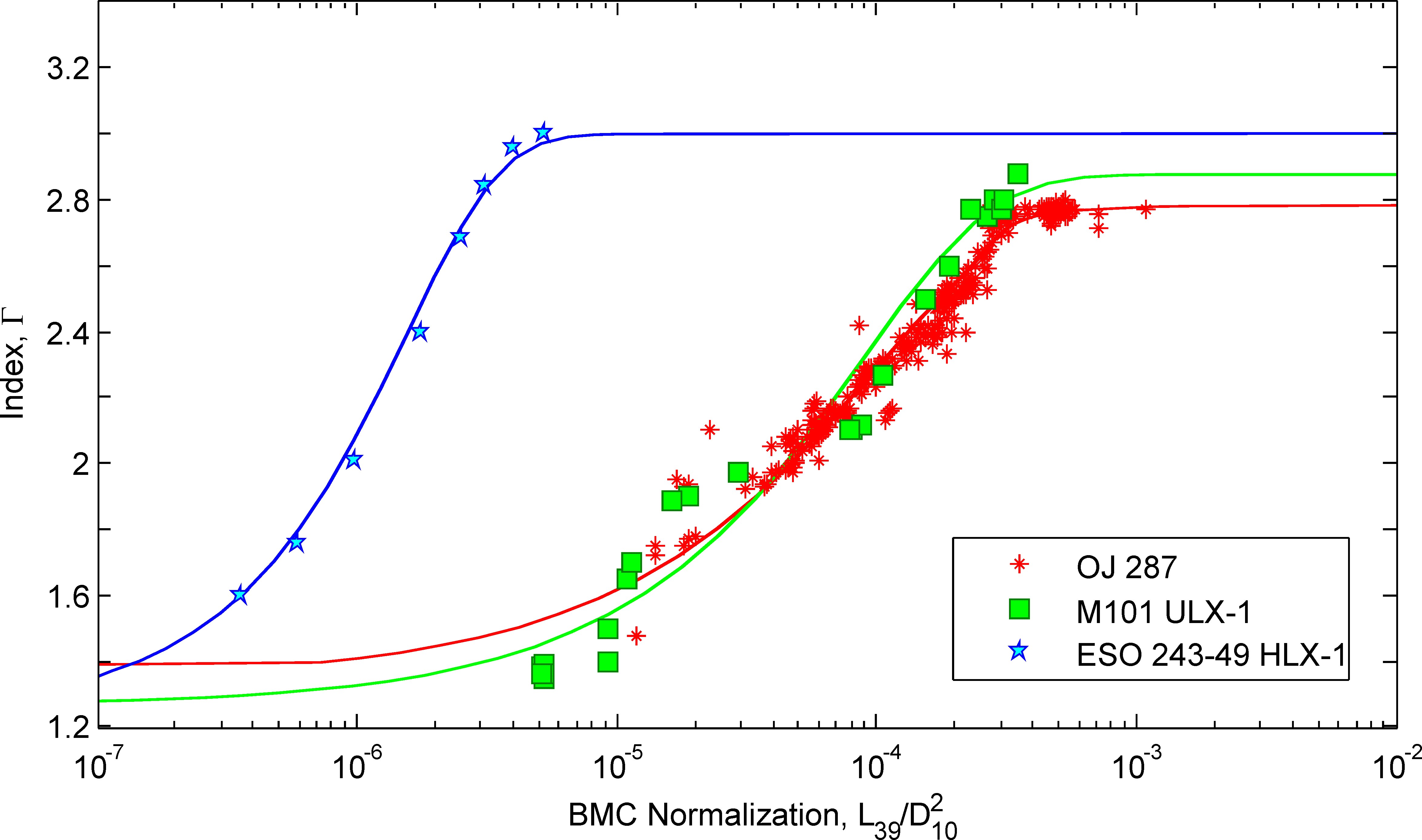}
      \caption{
Unique behavior of the photon index $\Gamma$ versus the normalization $N$ (proportional to mass accretion rate)  
for in the spectrum of OJ~287 during the outburst and its ``scaling” with an increase in the X-ray flux in units of L$_{39}$/D$^2_{10}$ (where $L_{39}$ is the luminosity in units of 10$^{39}$ erg/s and D$_{10}$ is the distance to the source in units of 10 kpc) for OJ~287 (red line) using data for extragalactic sources, ESO~243--49 HLX--1 and M101~ULX--1 (blue and green dots, respectively). Figure is taken from Titarchuk et.al. (2023).
}
\label{three_scal_1}
\end{figure*}

\begin{figure*}
\centering
\includegraphics[width=11cm]{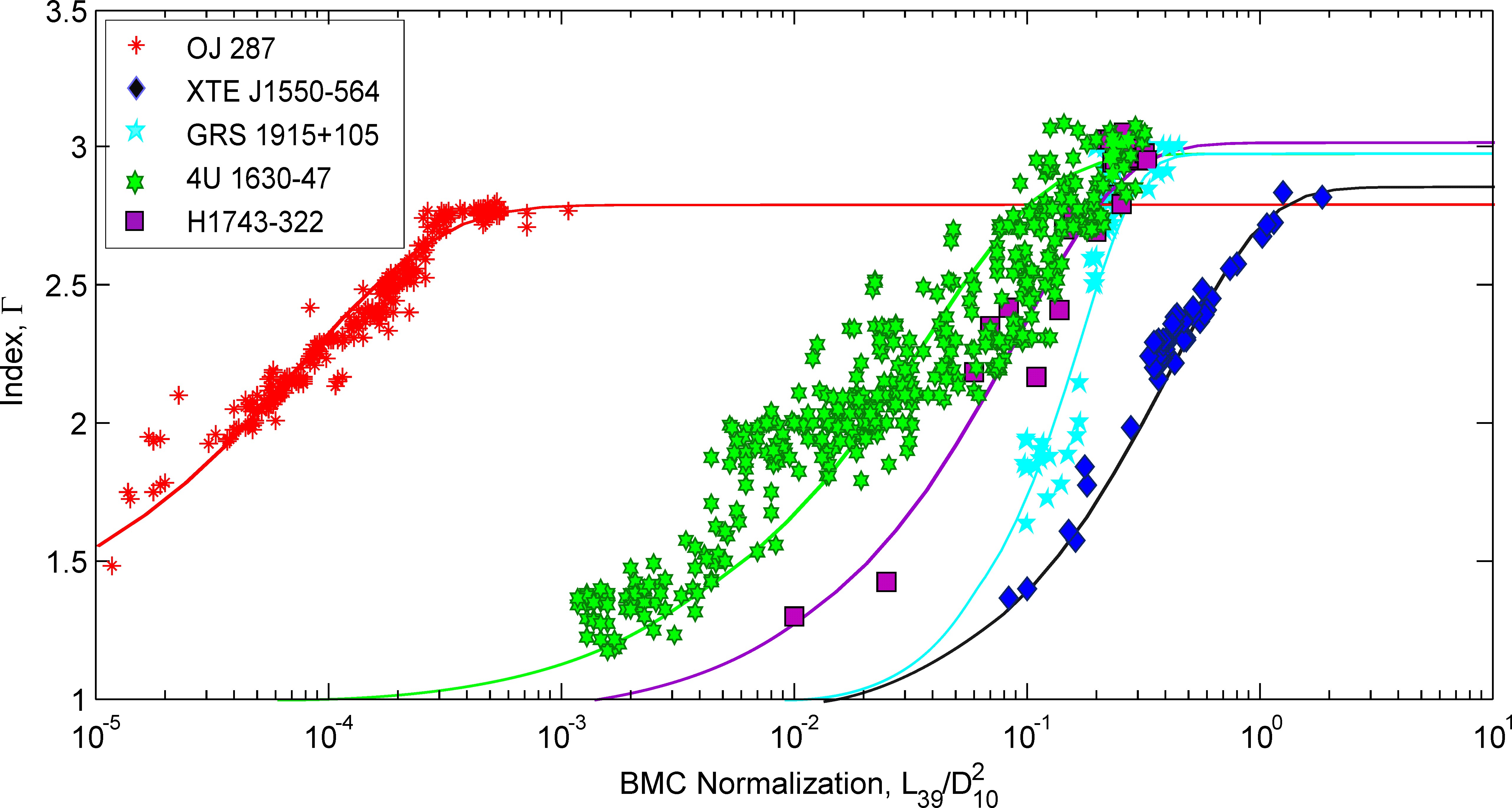} 
  \caption{Scaling of the photon index, $\Gamma$ versus the normalization $N_{BMC}$ for OJ~287 (red  points) 
using the correlation for the Galactic reference 
sources, XTE~J1550--564 (blue diamonds), H~1743--322 (pink squares), 4U~1630--47 (greed stars) and GRS~1915+105 (bright blue stars). Figure is taken from Titarchuk et.al. (2023).
} 
\label{three_scal}
\end{figure*}

It is known that such a unique behavior of $\Gamma$ is a spectral signature of the BH presence  in an object (Titarchuk \& Seifina, 2009). Based on this, Titarchuk et al. (2023) came to the conclusion that the ``culprit” of the X-ray outbursts in OJ~287 is an orbiting smaller black hole. To ``weigh” it, the ``scaling” procedure was used with extragalactic (Figure~\ref{three_scal_1}) and galactic (Figure~\ref{three_scal}) BH sources  as reference sources using ``scaling" technique (Seifina et al., 2014).  
In this way, Titarchuk et al. (2023) were able to determine the BH mass to be $\sim 1.25\times 10^8$ M$_{\odot}$, assuming a source distance of about 1~Gpc. 
This value is in excellent agreement with previous estimates (Valtonen et al., 2012).

Thus, it was possible to reveal the nature of X-ray outbursts in OJ~287 based on spectral analysis of X-ray emission. In this case, the latest achievements in modeling the BH states applying the Monte Carlo method were used based on the numerical solution of the complete relativistic kinetic equation and the detection of the ``saturation” phase of the spectral index during the peak of X-ray outburst in the BH source.

The method was developed and tested by us on a variety of astrophysical objects and showed excellent agreement with classical methods (
Seifina \& Titarchuk, 2010; Seifina et al., 2014; Titarchuk \& Seifina, 2016a,b, 2017; Seifina et al., 2018a,b; Titarchuk et al. 2020; Titarchuk \& Seifina, 2023; Titarchuk et al., 2023). The method is based on first principles and an assessment of the gravitational effect of a black hole on the matter around it, as well as calculating the effective size and mass of the response zone to such an impact. 

%
%

\subsubsection{
Discrepancy in the BH mass estimate from X-ray and optical data, as a possible indication on BBH in AGN center}
\label{pecularity} 
\begin{figure}
\centering
\includegraphics[scale=0.70,angle=0]{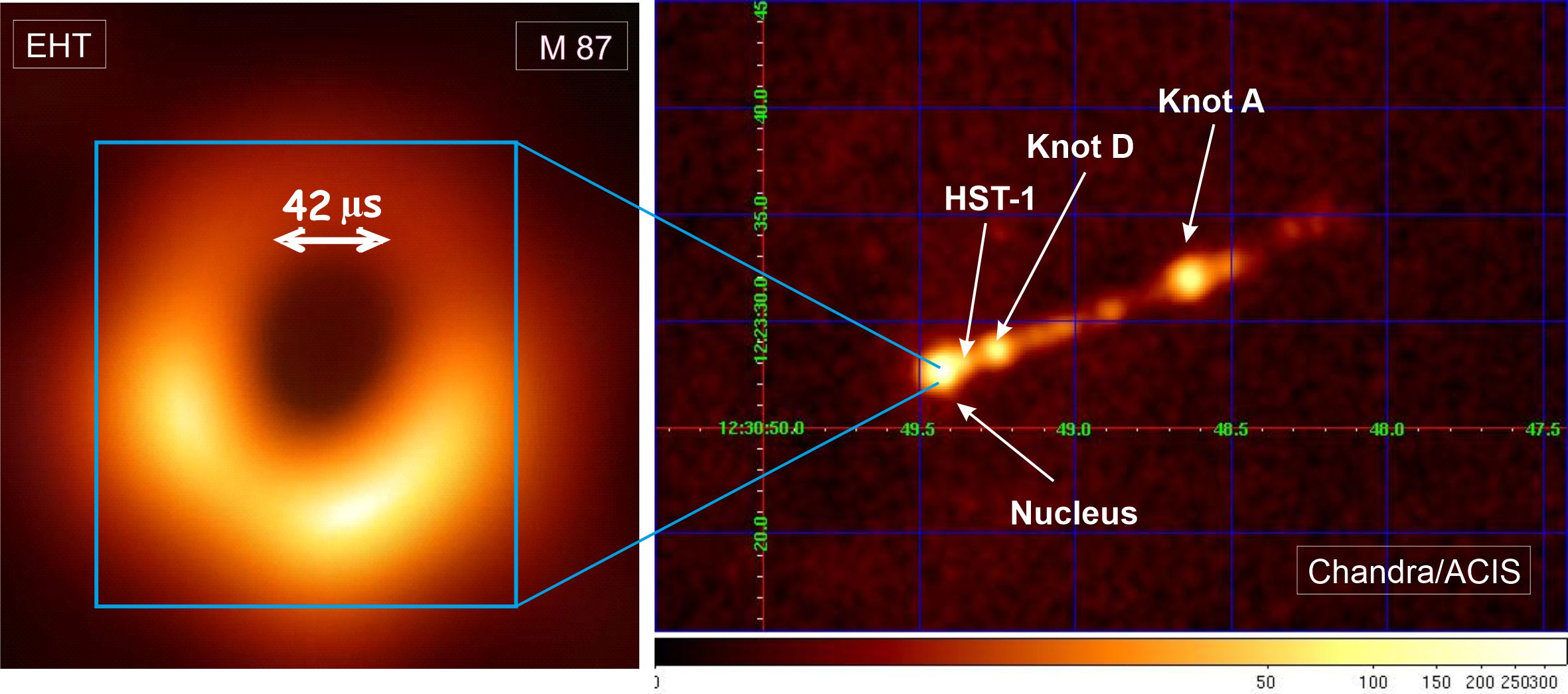}
\caption{Images of the black hole at the center of galaxy M87 with the {\it Event Horizon Telescope} (EHT, left) and {\it Chandra}/ACIS (right).  
The nucleus and jet knots (HST--1, D and A) of M87 are indicated in the right panel, while 
an enlarged image of the nucleus central part of M87 is presented in the blue box in the left panel. 
Right figure is taken from Titarchuk et.al. (2020).
}\label{fig:example_5}
\end{figure}

It is well known that the galaxy M87 hosts a supermassive black hole. It is suspected that this black hole may be a binary black hole. However, its duality is not so obvious and can be hidden in the complex structure of the central region polluted by the jet radiation (Figure~\ref{fig:example_5}).  Titarchuk et al. (2020) estimated its mass from {\it Chandra} X-ray observations using the ``scaling" method and compared it with the results of Stellar dynamics (Gebhardt et al., 2011; Akiyama et al., 2019) and an estimate from EHT radio observations (Akiyama et al.,  2019). Interestingly, the ``X-ray” results were very different from the others. Namely, according to X-ray data, the mass of the BH in M87 turned out to be 10$^7$ M$_{\odot}$, which is 100 times less than the mass estimates of this BH using other methods (10$^9$ M$_{\odot}$). The question arose why the mass of the black hole in (Titarchuk et al., 2020) turned out to be two orders of magnitude less? In view of eliminating all possible errors, the idea arose that such a difference could arise if Titarchuk et al. (2020) weighed a small black hole in a pair of two black holes, assuming that the BH at the center of M87 is a binary (rather than a single) BH. That is, the central BH in M87 Titarchuk et al (2020) was ``not seen” because it does not manifest itself in X-rays, and the small BH that flares up in X-rays is noticeable to an X-ray observer and therefore it was possible to weigh it using X-ray data. Thus, the difference between the ``optical” and ``X-ray” mass of a black hole may indicate its duality.

There is also an indication that the supermassive black hole in the center of our Galaxy is also binary BH (or was binary in the recent past and is now already at the post-merger stage). This is indicated by the atypical stellar population in one of its arms, which may be the remaining surviving ``island” of the galaxy that our Galaxy absorbed.

%
%

\section{Conclusions}

A review of current observational and theoretical researches of AGNs is presented along with brief historical perspective. The emergence of the binary BH hypothesis in its current form took almost 100 years. This hypothesis changed, modified, updated and supplemented along with new discoveries and observations. New results in this field are discussed, including the problem of close binary systems consisting of supermassive black holes in AGNs. 
Observational criteria for the duality of such black holes in AGNs are presented. It is shown that the detected binary BHs in AGNs fit well into the classification of already known AGN types, but their observational manifestations represent a new type of AGN variability.

The presented observations and analysis of the dual nature of BHs in AGNs confirm the reality of BH mergers in galactic nuclei, previously predicted by theoretical scenarios. Furthermore, the discovery of close binary systems hosted SMBHs solves the problem of the origin of ultramassive black holes ($>10^{10}$ M$_{\odot}$). In fact, such black holes could not form by accretion onto stellar-mass black holes  (3$\div$10 M$_{\odot}$). In fact, it would take longer than the lifetime of the Universe. Therefore, the discovery of binary AGN nuclei, observations of SMBH merger events and confirmation of the huge masses of such BHs obtained in recent years is a real breakthrough in understanding the origin of ultra massive BHs. 

The problems of SMBH ($\ge 10^6$ M$_{\odot}$) mergers  strongly overlap with mergers of lighter BHs, such as stellar-mass BHs. 
However, the latter can occur everywhere, and SMBH mergers are observed mainly in the nuclei of galaxies. In addition, mergers of stellar-mass and supermassive BHs have different spatial scales. 
Galaxies and their nuclei are huge structures and weighing them, in particular ``weighing" their binary nuclei, is a non-trivial task. Nevertheless, using the scaling method, it will be possible to determine the mass of objects in such systems.  Furthermore, it is always desirable to weigh in fundamentally different ways for the reliability of BH mass estimates. Note that previous methods were based on a variety of model assumptions. Therefore, weighing the black hole in OJ287 with a new method and its agreement with other estimates gives confidence in the correctness of the approach in general and the acquired knowledge about how the world around us works.

A rigorous analytical model of binary systems hosting SMBHs and their final merger in AGNs has not yet been created, so most of the results have so far been obtained using computer simulations. In the final stage of their orbital evolution, gravitational wave radiation provides the binary inspiral. 
It is gravitational waves that are responsible for the approach of SMBHs, taking with them the rotational energy. Black hole mergers cause a burst of gravitational waves and such events can only be detected using gravitational wave detectors such as LIGO and VIRGO. These instruments track how spacetime is ``stretched” under gravitational waves. The current era of gravitational wave astronomy has already provided observations of stellar-mass BH mergers, which are perfectly described using general relativity. However, to detect mergers of slowly approaching binary supermassive BH (with masses of 10$^8$ -- 10$^{10}$ M$_{\odot}$), gravitational wave detectors in the nanohertz range are needed. Therefore, great hopes are placed on pulsar networks and on the space laser interferometer eLISA, whose launch is tentatively planned for 2034 and which will operate in the nanohertz wavelength range and have sufficient sensitivity to detect these truly colossal emissions of gravitational energy.

In the context of the above, it can be assumed that other previously discovered galactic nuclei may actually be binary systems of supermassive black holes, and there may be more such black holes. The physics of AGN galaxies is the cutting edge of astrophysics, where the latest physical theories are used and tested, and there is great scope for new fundamental ideas, for the use of modern instruments and methods of observation and computer modeling. Perhaps future observations will show that SMBHs are much larger in the observable Universe, at least due to their duality in AGNs.

\section*{ACKNOWLEDGEMENTS}

The author expresses gratitude to A.D.~Kalinkin for his assistance in preparing the section on the AGN history. 
 The author also acknowledges the reviewer's comments, which improved the paper.

\end{document}